%% file: main.tex
\documentclass[a4paper]{ledger}
\usepackage{preamble}

\newcommand{\thefirstpagenum}[0]{1}

\hypersetup{pdfauthor={Vero Estrada-Gali\~nanes}, pdftitle={Towards Efficient Data Management For IPFS-based Applications}}

\title{Towards Efficient Data Management \\For IPFS-based Applications}

\author{Vero Estrada-Gali\~nanes\thanks{VEG (vero.estrada@epfl.ch) is Research Scientist at EPFL, Switzerland}\and Ahmad ElRouby\thanks{AE is Master Student at EPFL, Switzerland} \and Léo Marc-André Theytaz \thanks{LMAT is Master Student at EPFL, Switzerland}}

\pretitle{
  \centering \selectfont RESEARCH ARTICLE \par 
  \fontsize{24pt}{28pt}\selectfont} 

\begin{document}

\maketitle

\input{inputs/abstract}

\input{inputs/intro}
\input{inputs/ipfs}
\input{inputs/proposed_solution}
\input{inputs/AEcodes_ipfs}
\input{inputs/collaborative_repairs}
\input{inputs/evaluation}
\input{inputs/conclusion}


\ledgernotes

\section{Acknowledgement}
We thank Qiyuan Liang and Yuening Yang for their early contributions to the project.
This work was supported in part by the Initiative for Cryptocurrencies and Contracts (IC3) at Cornell Tech in partnership with Ethereum, Grant No. 10232.  








\input{biblio.bbl}
\appendix
\
\input{inputs/app1}
\input{inputs/app2}
\input{inputs/app3}
\input{inputs/app4}





\end{document}

%% file: inputs/abstract.tex
\begin{abstract}

Inefficient data management has been the Achilles heel of blockchain-based decentralized applications (dApps). 
An off-chain storage layer, which lies between the application and the blockchain layers, can improve space efficiency and data availability with erasure codes and decentralized maintenance.      
This paper presents two fundamental components of such storage layer designed and implemented for the IPFS network. 
The \textit{IPFS Community} is a component built on top of the IPFS network that encodes and decodes data before uploading to the network. 
Since data is encoded with alpha entanglement codes, the solution requires less storage space than the native IPFS solution which replicates data by pinning content with the IPFS Cluster. 
To detect and repair failures in a timely manner, we introduce the monitoring and repair component. 
This novel component is activated by any node and distributes the load of repairs among various nodes. 
These two components are implemented as pluggable modules, and can, therefore, be easily migrated to
other distributed file systems by adjusting the connector component.
\end{abstract}

%% file: inputs/intro.tex
\section{Introduction}

The emerging Web 3.0 (Web3) movement~\cite{ray2023web3} has the ambition of re-decentralizing web platforms and applications. 
Towards this end, many blockchain-based decentralized applications (dApps) and web3 technologies are designed with decentralized architectures that include a layer for distributed ledger technologies (DLTs).  
A caveat is that their blueprint usually omits or trivializes the storage layer and the components that are needed to manage data availability efficiently. 

Blockchains, a kind of DLT, need to be considered carefully for the need of reliably storing data and making it available to users. 
Replacing a traditional database for a blockchain makes sense to guarantee data integrity and content redundancy when mutually mistrusting entities need to interact and change the state of the system without using a third party\cite{wust2018you}. 
Even if blockchain technologies can be used in several use cases, they are not a feasible solution for several applications. 
One problem is that the cost of on-chain data is high and volatile. 
In Ethereum~\cite{buterin2014ethereum}, this cost is usually quantified in terms of the gas cost. 

The gas cost for on-chain data is only the tip of the iceberg. 
A typical blockchain-based solution requires and excessive amount of network resources such computation and memory. 
The number of content replicas cannot be controlled, any node that downloads the blockchain is storing a replica.
On plus, the more resources we asked from blockchain nodes, the more is the risk of centralization because a small operator may not be able to run a node that fulfill the requirements. 
Resource mismanagement can trigger further problems. 
For example, Cryptokitties~\cite{jiang2021cryptokitties}, now an outdated game application, at its peak of popularity has congested the Ethereum network and raised the gas price affecting the whole ecosystem. 

One way to address scalability and speed issues inherent in the original blockchain design (Layer 1) is to process transactions off the main blockchain as in Layer 2 solutions such as the Lightning network and rollups. 
Layer 2 solutions are designed to operate on top of a Layer 1 blockchain.
Another approach is to use a decentralized peer-to-peer network such as the InterPlanetary File System (IPFS)~\cite{benet2014ipfs} to store data off-chain.
For example, a large file can be stored in IPFS while alleviating the storage requirements in the blockchain by solely requiring on-chain storage of its IPFS content identifier (CID). 
When considering these solutions, the blueprint for a scalable decentralized architecture includes the application, the storage layer, and the L1/L2 blockchains.  

Examples of decentralized storage network layers are Arweave, IPFS, Storj, and Swarm. 
These networks are designed to store any kind of unstructured and semi-structured data (videos, images, JSON files, etc). 
Databases have been built on top to tolerate structure data and queries. 
The most relevant functionality of these networks is to add redundancy by storing the content in multiple nodes.
The replicas are added to increase the probability that the content is available when a user requests it. 
But n-way replication is not space efficient, even if the number of copies can be controlled by the network. 
On top, storing multiple replicas does not guarantee content availability.
At least one of the nodes responsible of the replicas need to be online but also the metadata to access that node needs to be available.
Any reliable system needs some failure detection mechanism and repairs to maintain file availability.
Inefficient data management has been the Achilles heel of dApps.

This paper presents self-organizing tools for encoding, decoding, and maintaining redundancy levels to facilitate efficient data management in a decentralized storage network.
Our work is built on top of the IPFS network, which is self-defined as an ``open system to manage data without a central server.''
Although IPFS introduces various components to manage data, it does not have any native solution for erasure coding data and maintenance. 
Instead IPFS relies on replication, that can be managed by third party pinning services or in a self-organized manner with the IPFS Cluster tool. 

The contributions of this work include: 
1) An investigation on multiple approaches to store encoded data and the rationale of using alpha entanglement (AE) codes. 
2) The design and development of a component built on top of the IPFS Cluster that encodes and decodes data while improving the space efficiency of the network (IPFS Community). 
And, 3) the design and development of a self-organized tool to repair missing files collaboratively.

%% file: inputs/ipfs.tex
\section{IPFS: A Primer on its Reliability}
\label{ipfs}

The InterPlanetary File System (IPFS) is a decentralized peer-to-peer file-sharing protocol, facilitating the distribution and retrieval of files within a network architecture devoid of centralized control mechanisms. This system is increasingly integral to dApps developments within the domain of Web3 technologies, offering a novel approach to data storage and access that diverges from traditional centralized data storage models. IPFS enhances the efficiency and speed of data sharing across a distributed network, thereby addressing significant challenges associated with data sovereignty and access latency~\cite{tschorsch2022ipfs}.

IPFS is conceptualized as a foundational infrastructure for Web3 applications, necessitating high levels of reliability and availability to support the operational demands of decentralized networks. 
Despite its innovative contributions to decentralized file storage and access, IPFS encounters several notable challenges that warrant further research and solutions. 
These challenges include a limited broader adoption, uncertainties regarding the protocol's scalability and performance under conditions of increased network load, concerns related to user accessibility and interface usability, and complexities associated with the upload process for large data files.

Some of the core components of IPFS relevant to this work are the distributed hash table (DHT), the Merkle Directed Acyclic Graph (DAG), and the pinning service provided by the IPFS Cluster. 
The DHT is a distributed key-value storage system, which provides IPFS with distributed storage. 
The Merkle DAG is a tree-like data structure used to ensure the integrity and verifiability of files stored in IPFS. 
Files added to IPFS are divided into blocks and put together in a Merkle DAG. 
The root hash is included in the content identifier (CID) which is used to access and route to the file via the DHT. 
Following this structure, files are immutable, and CIDs serve for integrity checks. 
By default, the file has only one replica that resides in the local IPFS node.
IPFS disseminates provider records upon file upload. 
These records facilitate the discovery process within the network, enabling other users to locate and retrieve files from the hosting node. 
To locate and announce content, IPFS uses the Kademlia DHT~\cite{maymounkov2002kademlia}.
Users may choose to become additional providers of a file, increasing its availability and resilience within the network, particularly for content that is in high demand or not supported by sufficient hosting resources.
The IPFS Cluster is a tool to create a private network of IPFS nodes that replicates content and interacts with the IPFS public network through proxies.

Some of the ongoing research and development efforts aim at enhancing protocol efficiency, user experience, and overall system scalability. 
Addressing these limitations is paramount to fully leveraging IPFS's potential in supporting the decentralized web and fostering the growth of dApps reliant on robust, decentralized data storage solutions.
In this section, we discuss the IPFS perspective with regard to reliability, the concept of pinning and the IPFS Cluster component. 
More insights about the IPFS monitoring metrics, failure detection, and the data allocation procedure are given in the Appendix~\cref{IPFS_insights}.  

\subsection{Reliability}
IPFS operates on a content-addressable storage model, meaning that content is retrieved based on its hash rather than its location. 
While this model enhances privacy and security, ensuring that content remains accessible without centralized control, it also introduces challenges in content persistence and availability. 
Specifically, if a file is not actively hosted by a node in the network, it may become inaccessible to other users. 
Unlike traditional web hosting, where a central server ensures content availability, IPFS relies on the network's nodes to voluntarily host and share content. 
This decentralized nature poses potential issues:

\begin{enumerate}
    \item Persistence: Content might become unavailable if no node chooses to host it or if the hosting nodes go offline. This is particularly problematic for less popular or older content that may not be actively accessed or maintained.
    \item Replication and Redundancy: Effective strategies for content replication and redundancy are necessary to prevent data loss and ensure high availability, which requires coordination and cooperation across the network.
    \item Incentivization: Mechanisms to incentivize nodes hosting and sharing content are required to ensure that data remains accessible over time. While solutions like Filecoin~\cite{benet2018filecoin} have emerged to address this by creating a market for storage, its effectiveness and integration within the broader IPFS ecosystem continue to be areas of active research.
\end{enumerate}

IPFS has implemented a design decision that, while advancing its efficiency and privacy objectives, introduces certain operational complexities. 
Specifically, newly uploaded files are initially stored exclusively on the uploader's local IPFS daemon (node).
This model protects user privacy by preventing the storage of data on potentially insecure nodes. 
The node operator has control over their node's content, offering a safeguard against the involuntary hosting of undesirable files. 

\subsection{Content Persistence via Pinning}
IPFS uses a garbage collection to automatically reclaim memory by eliminating content that is no longer used.
Node operators can choose to make content persistent by ``pinning'' it. 
Pinned content is not removed from the local cache during garbage collection. 
Although pinned content may be considered as ``reliably stored'', that is not a precise definition. 
If the local node is offline, the content will not be accessible.
Additionally, pinning can be done remotely with third-party services. 
The \textit{remote pinning} ensures that data persists on IPFS even when the local node goes offline or fails.
Another alternative is the IPFS Cluster, a self-organizing tool for content persistence among multiple IPFS nodes. 
Our work seeks to complement the functionalities enabled by the IPFS Cluster. 

\subsection{Replication}
IPFS does not enforces any replication mechanisms to protect content. 
As we discussed in the previous subsection, content that resides on a single node does not have protection against failures.
Also, IPFS lacks of a reliability mechanism to replace missing copies of a file without another peer accessing the data. 
There is no way for a peer in IPFS to improve its file's lifetime if nobody else downloads it, which limits the reliability offered by the system. 
This is where the IPFS Cluster can help.

\subsection{IPFS Cluster}
The IPFS Cluster enables collaboration between IPFS nodes and provides fault tolerance by allocating, replicating, and tracking a global pin-set distributed among multiple IPFS nodes. 
The pin replication is handled with the Raft consensus~\cite{ongaro2014search}. 

Let us break down the steps that peers must go through when pinning a file. 
Initially, the interested user will pin the file on its own IPFS node. 
The pinning operation takes an optional replication factor as an argument or takes the default configuration value. 
This number indicates how many peers must be selected (the desired total number of replicas in the network). 
Then, an allocation procedure (described in Appendix~\ref{alloc}) will locally produce a list of peers responsible for storing the file. 
The pinning initiator submits this list, along with other information, to the consensus component. 
Once the consensus validates this update, it integrates the global pinset and triggers a tracking operation at each peer.
This operation asks each peer to check if it is responsible of storing and tracking this file.
If so, the peer will pin the file on its IPFS node and start tracking it to react in case of failures. 
Afterwards, IPFS-Cluster continuously monitors its peers to maintain the appropriate replication factor even in the face of peer departures or failures.

%% file: inputs/proposed_solution.tex
\section{Proposed IPFS Community}

This section describes the high-level architecture of the proposed IPFS-based storage layer. 
This layer falls between dApps and L1/L2 blockchains.
It relies on the IPFS network and the IPFS Cluster to handle the uploading and downloading of data and parity blocks, but further enables the capabilities by maintaining state through a daemon running in the background and allowing collaboration between community nodes.
The collaborative aspect allows for faster and deeper repairs while monitoring improves block availability.

For clarity we describe our goals, a strawman and the improved proposal to align better with our goals. 
We also describe the rationale behind the chosen encoding algorithm. 

\subsection{Strawman}

Our main goal is to increase file reliability in IPFS without increasing storage overhead.
For that, we need to replace the replication mechanism with encoding data, particularly AE codes (in section~\ref{ae_codes} we explain why we choose these codes). 

We can now discuss a few approaches for achieving our goal: (i) use a remote pinning service, (ii) building on top of the IPFS Cluster, or (iii) reengineering IPFS.

Solution (i) introduces an extra burden on users, i.e., users have to set up and possibly pay for the service. Users need to trust the third party. Meanwhile, how service providers store the user-uploaded data is out of our control. Service providers are free to use any redundancy schema that they have at hand, and deliver it to the users. As a result, we consider this solution less optimal.

Solution (ii) turns the IPFS Cluster into our main entry point of interaction with the outside world. Our solution would send ``entangled blocks'' to the cluster and the cluster would allocate them to cluster member nodes. Members shall trust each other. 

Solution (iii) would provide a seamless user experience with IPFS taking care of the content entanglement. Independently of how the entanglement is done (file-wise entanglement or cross-file entanglement) this solution may require excessive reengineering work without knowing if the changes would be accepted by the IPFS core team. 

As a compromise, we seek for solution (ii), which uses IPFS Cluster as the underlying mechanism to distribute entangled blocks. It is more flexible than solution (i) since we can tailor our own logic on top, and it is more feasible than reengineering IPFS.

\begin{figure}[ht]
\centering
\includegraphics[width=5in]{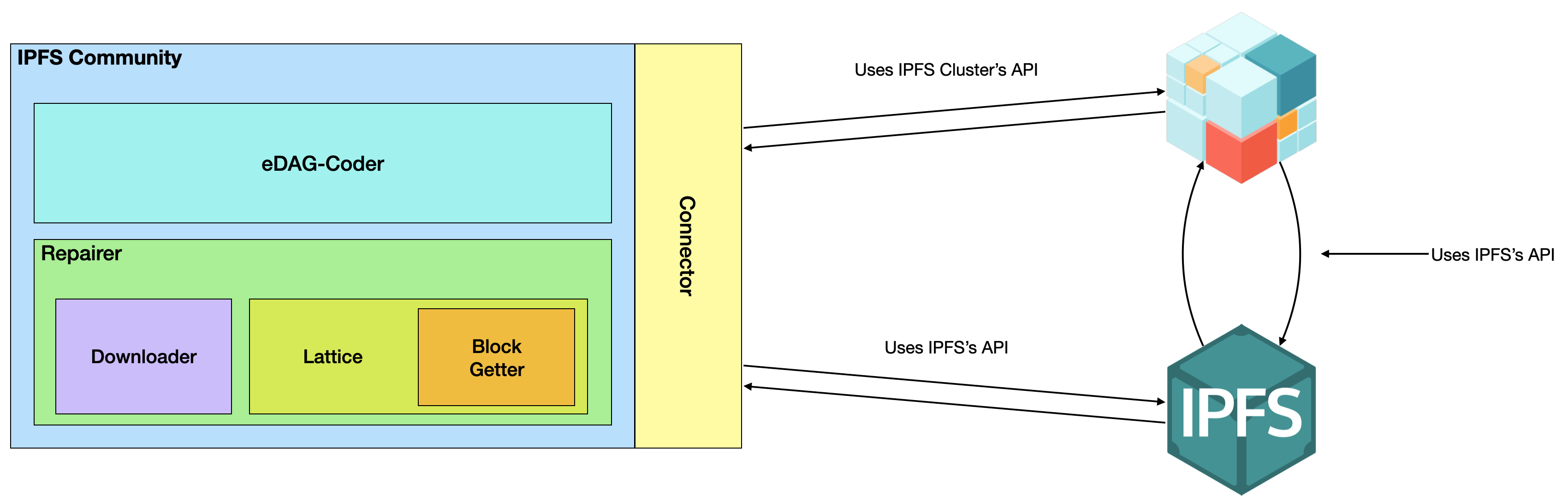}
\caption{Strawman IPFS Community Architecture. }
\label{fig:1}
\end{figure}

To introduce the design of the IPFS Community, we begin with a simple strawman illustrated in~\cref{fig:1} and described below. 
An IPFS Cluster node runs on top of an IPFS node. 
IPFS Cluster nodes themselves do not store the file content; it instructs the IPFS node to do so. 
The connection between different IPFS Cluster nodes is only privately known. 
On the other hand, IPFS nodes are all publicly available, meaning that contents stored in them could be accessed by others. 
Unless the node operator limits the accessibility of the IPFS nodes.

To simplify the implementation of this idea, this strawman uses a \textit{metafile}. 
The metafile contains information related to the encoding/decoding process. In the native IPFS solution, the user only provides the file CID to download a file. 
To be able to profit of the additional redundancy added with the encoded data, our strawman requires that the user provides the metafile as well as the CID. 
If the original files are missing, the download procedure will try to fetch the entanglement and restore the original file.

The IPFS Community has three major components, eDAG-Coder, Repairer, and Connector, each of them responsible for different functionalities. 
The user interacts with the IPFS community with a CLI interface.
When we upload a file into IPFS, a Merkle DAG is created as discussed in Section~\ref{ipfs}. 
The \textit{eDAG-Coder} takes the DAG as input and entangles its block components to create $\alpha$ entangled DAGs (eDAGs). 
The encoding parameter $\alpha$ is specified by the user to indicate the desired redundancy factor. 
The \textit{repairer} component is a complement to the downloader. If some or all blocks of a file are missing, the repair process attempts to recover the file using the entangled DAGs.
We assume both one IPFS Cluster instance and one IPFS instance run locally. 
The IPFS Cluster nodes interact with IPFS using its exposed APIs. 
The IPFS Cluster instance connects to cluster member nodes, while the IPFS instance connects to the IPFS public network. 
The Connector connects the IPFS Community to the running IPFS Cluster and IPFS instances. 
The IPFS Community uses the IPFS Cluster and the IPFS node as tools to deliver the service; it does not change their internal logic.

The following steps occur when uploading a file: 
(1) A user uploads a file through the IPFS community. The file is handled by the local IPFS to create a Merkle DAG where the leaves are the data blocks that form the file. 
(2) The eDAG-Coder retrieves all the blocks back to entangled them together, generating entangled blocks. 
(3) The eDAG-Coder uploads the entangled blocks to the local IPFS, making them locally available. 
(4) The eDAG-Coder pins the entangled content via the IPFS Cluster's API, making the blocks remotely available. 
(5) The user receives the file CID and the metafile CID.

The following steps occur when downloading a file using: 
(1) A user input the file's CID and the metafile CID. The metafile CID is optional but useful for repair missing data. 
(2) When doing repairs, the eDAG-Coder triggers the building of a lattice (related to the AE code algorithm) based on the information provided in the metafile, e.g., the total number of data blocks. 
(3) The lattice requests any additional desired block from the IPFS network. 
(4) The Repairer repairs missing blocks using the additional blocks. 
(5) The user obtains the requested file.

\subsection{Challenges}
The metafile of our strawman, from now on also referred as \emph{IPFS-community 0.5v}, contains all the entanglement configuration parameters and file information relevant to the Repairer, as well as a mapping between the lattices indexes and the CIDs for data/entangled blocks. 
This approach introduces an additional complexity for the user, who needs to keep the metafile, limits the repairs to only users who have the metafile, and requires additional space overhead (something that we want to avoid).  
According to our estimations (see Appendix~\cref{metafile_overhead}) for 300 nodes, 1000 files that generate 200 blocks each, the metafile generated when the encoding parameter $\alpha = 3$ would require approx. 9 GB.
This space requirement would increase further, the more nodes and files were added to the system. 
The takeaway is that our final design must not use a metafile. 

\subsection{Design Overview}

Our starting point is the IPFS Community v0.5 strawman. 
To add capabilities, the IPFS Community requires a stateful design. 
A daemon running in the background allows collaboration between community nodes and performs all operations on behalf of the user. 
Collaborations contribute to efficient and deeper repairs (using locality for single repairs but expanding repairs in cases with large amount of failures). 
The daemon continuously monitor the status of the files and a failure detection mechanism that triggers preventive repairs when detecting important loss of blocks. 

As a result, our proposed solution, from now on referred as \textit{IPFS community 1.0v} can profit from multiple peers collaborating in a file repair processes, which balances network traffic and speeds up the recovery of blocks (see Section \ref{performance}). 
The storage overhead caused by the metafile was reduced by creating our key data structure, the $\alpha$ entangled Merkle DAGs (see \ref{eDAG}). 

\begin{figure} [ht]
\vspace{10pt}
    \centering
    \includegraphics[width=3.5in]{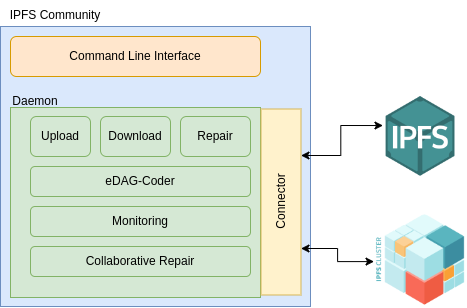}
    \caption{Design Architecture: IPFS Community Node}
    \label{arch}
\vspace{5pt}
\end{figure}

As outlined in figure \ref{arch}, the architecture of a single \emph{IPFS community 1.0v} node consists of the following: 
(1) \textit{Command Line Interface:} An interface that provides the user with the ability to upload, download, repair files as well as starting a daemon in the background to perform all the required operations. 
(2) \textit{IPFS:} Each community node relies on an IPFS peer node running on the same machine to which it forwards the requests for uploading and downloading blocks from the network.
(3) \textit{IPFS Cluster:} This component ensures that given at least one copy of a pinned block, it will be replicated according to the specified configurations. It maintains a pinset and runs the consensus algorithm to make sure all nodes have the same view of the pinset.
(4) \textit{Connector:} It maintains connections with the IPFS and IPFS cluster nodes. It also handles all communication required to perform any operation that involves either of these two components.
(5) \textit{eDAG-Code:} It uses AE codes to generate entangled blocks taking as input data blocks of a specific file.
(6) \textit{Upload:} It handles the file uploading process to IPFS, generating the entangled blocks through the eDAG-Coder, and creating entangled Merkle DAGs use for the repair mechanism.
(7) \textit{Download:} It requests data from IPFS, and subsequently returns the file to the requesting user in case of success, or starts a limited repair process in case of missing data blocks.
(8) \textit{Repair:} Essentially, it orchestrates all needed steps to successfully recover a file given the possibility and allowed repair depth (more about this later).
(9) \textit{Monitoring:} It runs in the background. This component monitors files under its responsibility to determine their status and triggers repair when deemed necessary.
(10) \textit{Collaborative Repair:} It coordinates between different peers to distribute the load for a deep repair of a certain file. 

\subsection{Key Algorithm: AE Codes}\label{ae_codes}
Our proposed solution relies on AE codes~\cite{estrada2018alpha}.
For more detail on how the AE codes work we refer the reader to the original publication. 
In this section we will elaborate on the decision of using AE Codes and address the following questions: (1) Why not replication? (2) Why not traditional erasure codes?

Distributed systems can increase reliability and fault tolerance by storing data replicas across multiple storage devices or locations. 
Redundancy prevents errors when a component is malfunctioning, there are transmission errors, or the node is not available. 
But using replication for decentralized solutions is not scalable in the long term. 
As a method to create redundancy, n-way replication is not space efficient. 
In addition, popular content triggers the creation of an excessive amount of replicas, congesting the network, and filling up memory space while leaving unpopular or ignored content without sufficient protection.  

In a trustless distributed environment, the mechanism that guarantees data integrity is fundamental. 
While the correctness of replicas can be verified with a quorum consensus mechanism in a distributed system, off-chain storage systems provide other options. 
In the context of IPFS, data integrity refers to the assurance that data has not been tampered with or altered in an unauthorized manner from the time it was published to the time it is accessed. 
To retrieve data, IPFS relies on content addressing, a method where each piece of content is given a unique CID that contains a cryptographic hash of the content itself. 
The use of CIDs ensures that references to data are immutable and independent of their location. 
If content is changed or updated, it receives a new CID, preserving the integrity of the original data. 
This means that the CID changes if even a small part of the content is altered, ensuring that any modification to the data results in a different identifier.
This prevents unauthorized alterations since any change to the content would be easily detectable due to the change in its CID.

CIDs can be verified directly by a user instead of the need to trust a consensus committee. 
When data is retrieved using a CID, the IPFS node can verify that the content matches the request by computing the cryptographic hash of the received content and comparing it to the requested CID. 
If the computed hash and the CID match, it confirms that the content has not been altered, ensuring its integrity. 
The cryptographic nature of CIDs makes it computationally infeasible to reverse-engineer the content from the CID or to create content that would match a specific CID. This property is essential for preventing tampering, as any attempt to alter the content would be easily detectable by anyone verifying the data against its CID.

The Merkle DAG structure allows each node to link to other nodes using their CIDs. Because these links are based on the content’s hash, they are immutable — altering the content at any point in the DAG would require changing all subsequent CIDs, making unauthorized alterations easily detectable. Complex data structures, such as files or directories, can be represented as a hierarchy of Merkle DAG nodes.

AE codes add extra layer for integrity protection with the created interdependencies between data and entangled blocks. 
The decoder algorithm can use different blocks to generate the same output. 
If that output is different, it becomes evident that something went wrong. 

Traditional erasure codes, such as Reed Solomon, does not have this additional layer for data integrity protection. 
When a file is distributed among network nodes, repairing missing chunks with erasure code requires more bandwidth. 
The problem is particularly relevant when repairing single failures.
Thus, a decentralized repair mechanism does not make any difference. 
On the contrary, the node in charge of repairs need to get all the needed blocks from the network and recover the file locally. 
While this can work for streaming applications, it is not a good option for random access and decentralized monitoring and maintenance of the storage network layer. 

\subsection{Key Data Structure: Entangled Merkle DAGs}\label{eDAG}
In our strawman, entangled blocks were uploaded to the IPFS network as single files. 
This individual entangled blocks were worthless without the information contained in the metafile. 
To enable seamless repairs, our solution embeds this information inside the IPFS hash tree, also known as Merkle DAG. 
Merkle Trees play a crucial role in how data is structured in modern decentralized storage systems to enable efficient content verification. 

Thus, we replace the metafile for a more efficient data structure. The entanglement process will generate a Merkle DAG for all the relevant entangled blocks, aka \textit{eDAG}, to a particular ``strand direction.''
A strand and the strand direction are concepts related to the AE codes. 
The parameter $\alpha$ indicates the total number of strands directions.
By creating $\alpha$ eDAGs, the system has multiple ways of repairing the file. 
First, the repair process can use any of the $\alpha$ eDAGs, since each of them is self-contained and can be used to recover the file by sequentially xoring the entangled blocks that form the chosen eDAG.
Alternatively, multiple eDAGs can be used in combination to repair a large number of missing blocks. 
Entangled Merkle Trees were introduced by~\cite{nygaard2021snarl}.  
In this work, we adapted the structure to the specifics of the IPFS Merkle DAG.

%% file: inputs/AEcodes_ipfs.tex
\section{IPFS Community Implementation}
\label{AEcodes}

The project was implemented during two semesters (28 weeks). 
The codebase has integrated more than 14k lines of code written in Golang, NodeJS, and Python and is open-sourced~\footnote{\url{https://github.com/dedis/student_23_ipfs_reliability}}. 

The starting point is the command line interface (cli), with its main command:

\begin{lstlisting}
$ daemon --cluster-ip <ci> --cluster-port <cp> 
--community-ip <ip> --port <p> --discovery <server> 
--ipfs-ip <ip_ipfs> --ipfs-port <port_ipfs>
\end{lstlisting}
is responsible for running the daemon that is used for multiple purposes as described earlier. The command requires information about the ip address and the port on which the daemon would run. Information regarding the IPFS and IPFS cluster nodes are also required. Finally the full address of the discovery address should be provided.

\subsection{Upload: File Entanglement}
The file entanglement is done during the file upload process initiated through the command line interface according to encoding parameters specified by the user. 
The output of the process are the Merkle DAG for the data file and $\alpha$ Merkle DAGs constructed with the entangled blocks. 
\begin{lstlisting}[breaklines]
$ upload --address <community_address:port> --alpha <a_AE> 
--p <p_AE> --s <s_AE> --direct-replication <replication_factor>
  --replication <rf_intermediate_nodes_EMTs>
\end{lstlisting}

\subsection{Download: File Repairs}
Another integral operation provided is the ability to download and perform limited repairs to recover a file. 
The main goal of this operation is to provide the user with either the recovered file or an error with minimal perceived latency. 

The successfully recovery of the file in this case is limited by the \textit{repair depth} specified by the user. 
The depth refers to how far we use a particular eDAG to recover a missing block before giving up (it indicates how many entangled blocks are recursively used to repair a file). 
After this step, we either have the complete file or not. 
Repaired blocks are uploaded to IPFS again. 
The download service returns the download data to the original requester if recovery was possible or an error otherwise.

\begin{lstlisting}[breaklines]
$ download --address <ip:port_community_node> --depth <depth>
  --metacid <metadata_CID> --output <path_downloaded_file>
  --upload-recovery <bool_upload_recovered_blocks_back_to_IPFS>
\end{lstlisting}

%% file: inputs/collaborative_repairs.tex
\section{Collaborative Repairs}

The goal of collaborative repairs is to distribute the load of maintaining file redundancy among peers. 
The benefits are to speed up the process and require less computer resources per peer. 
Additionally, we can also increase the depth of the repair, i.e., more entangled blocks that can be used to repair a particular failure, improving the utilization of the existing redundancy.

\begin{figure} [ht]
\vspace{10pt}
    \centering
    \includegraphics[width=4in]{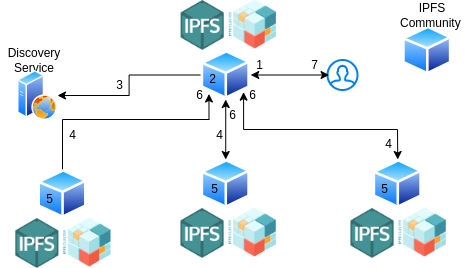}
    \caption{Collaborative Repair}
    \label{collab_repair}
\vspace{5pt}
\end{figure}

In this section, we describe how we managed to attain this goal. The figure \ref{collab_repair} shows an overview of the flow for running collaborative repair. The blue cubes in it represent the \emph{IPFS community 1.0v} node we built which in our case should run alongside an IPFS, and an IPFS-Cluster node on the same machine. It is important however to note that they don't necessarily have to run on the machine in which case the discovery service was deemed necessary.
The discovery service allows the community nodes to query the peers available on the private network and provides mapping from cluster nodes to community nodes. The discovery service runs in the background with a fixed hostname that is used by community nodes on startup to advertise their addresses. This solution was motivated by the timing constraints of the project. In order to relate IPFS-Community and IPFS-Cluster nodes together as well as benefit from the existing network parts of IPFS-Cluster in a short amount of time, we centralized the discovery service. It then performs regular health checks\footnote{These checks are solely for the purpose of delivering correct network addresses, and not related to data monitoring.} according to a pre-defined interval to make sure that its view of the system is correct. 

The repair flow includes seven steps. 
\begin{enumerate*}
    \item The collaborative repair in this diagram was triggered by a user for demonstrative purposes only. However, in reality, the repair is triggered by the failure detection service running in the Community node, or a failed download for a particular file where the repair with limited depth was not enough. 
    \item The node on which the collaborative repair was triggered should now be referred to as the coordinator. The coordinator starts by fetching the metadata from the IPFS network. It then starts to recursively traverse the original merkle tree of the file. For every missing non-leaf tree node, the coordinator attempts to recover in the background with a strand depth pre-defined in the system's configuration. In case of failure, the repair is aborted, and the system reports the operation a failure. However, if all intermediate tree nodes were recovered successfully, the coordinator will then try to fetch all leaf nodes. All leaf nodes that were unavailable will then be added to a list. If all node are available, then the coordinator reports the success of recovery process and returns the file.
    \item The coordinator then queries the discovery service for available peers. A peer-to-peer membership algorithm would have been required without the discovery service.
    \item The coordinator then chooses a specific number of peers (which is passed as an argument or pre-defined in the configuration) for the collaborative repair process, splits the unavailable leaf nodes amongst them, and then forwards them to these peers.
    \item Each peer now with a list of nodes (or data blocks) missing from the network, starts their recovery process. Individually, each peer runs multiple instances of recovery for the missing nodes and awaits the result. 
    \item The final result for each missing node (or data block) is then packaged as a response and sent back to the coordinator, where it re-assembles the file if the process was successful. 
    \item Finally, the coordinator asynchronously sends back the final state of the process to the requesting service. 
\end{enumerate*}

%% file: inputs/evaluation.tex
\section{Evaluation}

We evaluate the IPFS Community with a detailed experimental evaluation. 
We use a network with  20 IPFS nodes, IPFS cluster, IPFS community nodes.
All experiments were repeated 10 times and use AE codes with coding parameters: $\alpha$=3, s=5, p=5. 
The default fize size for uploads/downloads is 25 MB. 
\subsection{Environment Setup}
All experiments were carried out on an Ubuntu 22.04 running on a 16-core AMD Ryzen 9 5950X 3.4 GHz with a total of 64 Gigabytes of RAM. 
This setting allows to run multiple instances, one for each peer, on a single machine. 
We use Jinja2 library to generate different configurations for testing. 
We rely on docker compose to handle the creation of peer instances and managing the networking interfaces between them. 

\subsection{Recovering Data: Replication vs eDAGs}
\begin{figure}[ht]
    \centering
        \includegraphics[width=5in]{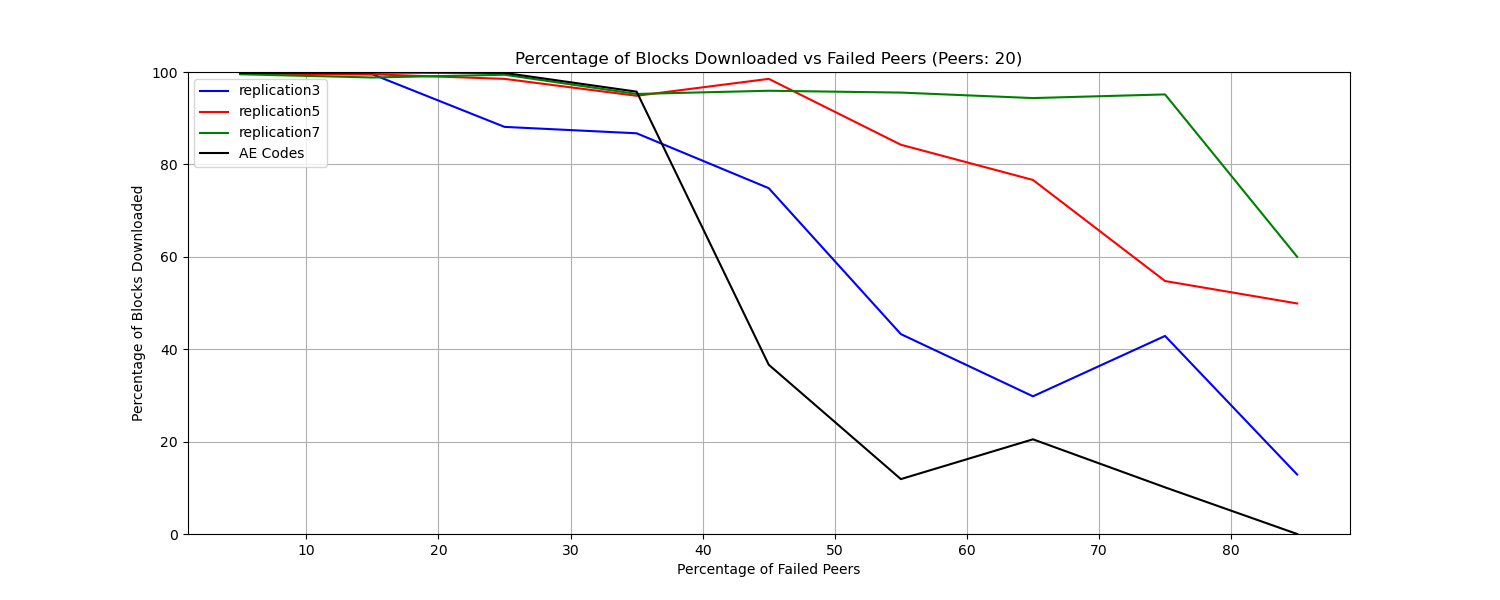}
        \caption{Average percentage of recovered blocks given a percentage of peer failure}
        \label{Percentage Downloaded 20}
\end{figure}

We run experiments with the strawman design to compare the recovery rates of AE codes with replication. 
Results are not included due to space reasons and are similar to those found in the literature~\cite{nygaard2021snarl}.
Contrary to the strawman design, the eDAGs that replace the metafile introduce another failure domain, in which a missing intermediate node in the Merkle tree would completely hinder the recovery process. 
Those nodes are protected with additional replication specified by the user.
We run experiments to compare the likelihood of recovering data either partially or fully using replication (R=3,5,7) and eDAGs implemented in the IPFS Community.
Results are presented in Figure~\ref{Percentage Downloaded 20}. 

\subsection{Performance Evaluation for Collaborative Repairs}
\label{performance}
We compare performance of collaborative repairs against single peer repair using identical parameters. 
Single repair total time refers to the total time taken to recover the file, and on the collaborative repair's side, the total time taken from finding which blocks are missing, distributing the failures across the peers, waiting for all peers to finish and finally reporting the success of the operation. 
End-to-end the procedure is the same since the goal was to determine how long it would take to finish the repair process for a single file. 
In addition, we compare the average processing time for each peer which is expected be lower in the collaborative mode since each peer is only repairing a fraction of the file. 
Finally, to determine the bandwidth overhead of both methods, we compare firstly the total number of blocks downloaded over the network for each method, as well as the average blocks downloaded per peer. 

 Measurements for the impact of repair depth and more figures appear in the Appendix~\cref{performance_insight}. The Appendix also includes measurements for the impact of repair depth. 
We summarize here the key takeaways:

\begin{enumerate}
    \item On average, total repair time in collaborative repair is significantly lower than its counterpart in single repair, especially with higher failure rates.
    \item On average, each peer would spend considerably less time repairing the file on the network compared to a node performing the repair alone.
    \item The average number of blocks downloaded over the network per peer is much lower in the case of collaborative repair over single repair due to the fact that collaborating peers only need to repair a fraction of the data. 
    \item On average, the total bandwidth used over the network downloading blocks is much higher in the case of collaborative repair. This can be attributed to the fact that a single node doesn't need to download the same block twice, however, multiple peers attempting to repair a single file will have overlapping blocks through the repair process, forcing them to download the same block multiple times. 
\end{enumerate}

%% file: inputs/conclusion.tex
\section{Conclusion}

Merkle DAGs are a core concept for decentralized systems, not only for storage networks like IPFS, but also at the core of technologies like git, bitcoin and dat. 
This work explores resource efficient data management methods for off-chain data using entangled Merkle DAGs and collaborative repairs.
Our results demonstrate significant space savings and a promising failure detection and repair mechanism to persist content in decentralized networks, thereby laying a solid foundation for future research and developments.
These advancements not only contribute to IPFS but also could benefit a wide range of decentralized applications.

%% file: inputs/app1.tex
\section{More Insights about IPFS}
\label{IPFS_insights}

\subsection{Monitoring}
In order to detect failures, peers must monitor each other. 
IPFS-Cluster peers do so using a Pub-Sub system within their \emph{monitor} component. 
A common topic is set up to share metrics with each other at regular intervals. 
Each metric has a TTL, and fresh metrics will be sent after half of the smallest TTL duration (among all shared metric types) using a timer based on the OS clock. 
For example, the TTL for a "freespace" metric is 30 seconds by default.

\subsection{Metrics}
\label{metrics}

Metrics are used to share information about peers. Those have a type, a value, an expiration date (in Unix time), a weight, and a "partitionable" attribute. It allows peers to handle the information they receive in the correct way. The metrics used in IPFS-Cluster have the following types:

\vspace{10pt}
\begin{itemize}
    \item freespace/reposize: Indicates how much free space the IPFS node has/the used space reported by IPFS.
    \item numpin: Indicates the number of items this peer is pinning.
    \item pinqueue: Indicates the number of pending pins at this peer.
    \item tag:tag\_name: Can indicate anything to help declustering the allocation process. For example, metric name "tag:region" with value "Florida" would indicate that the peer's server is located in Florida. The administrator of the IPFS-Cluster can set the tag name and values in each peer's configuration.
\end{itemize}
\vspace{10pt}

The partitionable attribute optimizes the allocation process by only allowing meaningful partitions. When selecting peers to replicate a file, it makes sense to spread the storage as much as possible. 
Targeting peers with different characteristics (such as a "tag:region" for example) will declusterize the allocation to improve reliability. 
Some metrics give insights of (de-)correlation between peers, and others help to rank peers based on their capabilities. 
For example, \emph{freespace} is a non-partitionable metric, and will only be used to sort peers and pick the ones with the most freespace available.

\subsection{Allocation procedure}\label{alloc}
The allocation procedure seeks to distribute data across the best available peers, reducing the risk of single point of failure that would cause data loss. 
This procedure outputs the list of peers across the cluster that will pin the file on their IPFS node. 
To decide which ones will be selected, the node takes multiple variables into account: 

\vspace{10pt}
\begin{itemize}
    \item min/max replication factors (default or user-selected)
    \item the user-set allocations (overriding the allocation procedure)
    \item metrics about the other peers based on the configuration of the allocator (\emph{allocate\_by})
    \item potential existing allocations (if the CID was already pinned).
\end{itemize}
\vspace{10pt}

\subsection{Failure Detection}
\label{decay}
Careful allocation and monitoring contributes to prevent file loss and that the system degrades. 
The risk of file loss increases with every replica that becomes unavailable.
When a peer does not receive an up-to-date metric before the metric expiration time\footnote{IPFS-Cluster use OS clock time and assumes negligible clock drift for the timeouts to be fixed.}, it raises an alert. 
Peers will then verify in the global pin-set if the node they suspect to have failed was in the allocation list of a pin. 
If so, the closest alive peer to the failed peer will trigger an operation to re-pin the file.

In case every peer which had pinned this file were to disconnect at the same time, the re-pin operation would fail, as the original data would be nowhere to be found. 
More replicas reduce this risk by increasing reliability at a high cost.
Given the constraints of limited storage capacity, exploring alternative methods to enhance reliability presents a compelling opportunity.

%% file: inputs/app2.tex
\section{Failure Detection in the IPFS Community}
\label{app2_monitoring}

IPFS-Community offers storage reliability with redundancy in the format of AE codes to reduce the storage overhead of simply replicating a file multiple times. To do so, parities resulting of the entanglements and data blocks are pinned at the IPFS-Cluster layer with a replication factor of 1. While the blocks are tracked, there is no guaranteed way to recover such blocks if a peer goes offline. The repair mechanism using AE codes allows for missing blocks while still recovering the original file, but the maximum number of missing blocks tolerated depends on the AE parameters $\alpha$, $s$ and $p$. Our goal is thus to detect those losses before the file becomes irrecoverable and trigger preventive repairs in batches to optimize the repair costs.

\subsection{Monitoring nodes}
\label{monitoring_nodes}
Before triggering preventive repairs, block losses have to be detected. To achieve this goal, some peers must actively monitor the network and look after failures. In order to select these \emph{monitor nodes}, we will reuse the allocation process of IPFS-Cluster. Indeed, when a file is added through IPFS-Community, root nodes of the different parity trees will be pinned to IPFS-Cluster. If monitoring is requested for a file, the peers selected during the allocation process to pin the strand roots will be designated as \emph{monitor nodes}. In Figure \ref{hl_monitoring} we can see how the monitoring is put in place:

\begin{figure}[ht]
\vspace{10pt}
    \centering
    \includegraphics[width=4in]{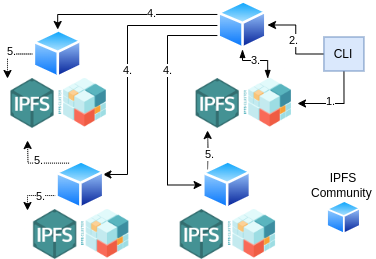}
    \caption{High level monitoring set-up}
    \label{hl_monitoring}
\vspace{5pt}
\end{figure}

\begin{enumerate}
    \item The command line interface uploads the file to IPFS and parities to IPFS-Cluster which results in allocation processes.
    \item If monitoring is requested, the CLI will ask a Community node to initiate the monitoring.
    \item The Community node initiating the monitoring gets the allocation list for each strand root of the entangled file.
    \item The initiator sends a request to start the monitoring to the \emph{monitor nodes} present in the allocation list.
    \item Finally, the \emph{monitor nodes} can start checking for the presence of blocks for this file and trigger repairs when deemed necessary.
\end{enumerate}

The implementation of \emph{IPFS community 0.5v} is a command line tool that interacts with IPFS and IPFS-Cluster nodes. To allow for active monitoring, in \emph{IPFS community 1.0v} we add a community node on top of each IPFS-Cluster node. After uploading a file with \emph{IPFS community 1.0v}, peers selected to track strands' root will get a request at their Community node to start looking after missing blocks. Each \emph{monitor node} will be tasked to: 1. Get a view of the cluster and files (failed regions, failed blocks, time between block failures, ...), 2. Heuristically check that blocks of the monitored files are present, 3. Trigger preventive repairs if a criticality threshold is crossed, and 4. Sharing views and restarting\footnote{Assuming that a crash of the Community node implies a crash of the corresponding Cluster node, a re-pinning operation will be triggered if the Cluster node was responsible for pinning blocks. If it pinned a strand root, it means we have to restart monitoring at another Community node.} monitoring upon failure.

%% file: inputs/app3.tex
\section{Metafile Overhead Calculations}
\label{metafile_overhead}
To estimate the additional storage related to the metafile we use the following equation:

\begin{equation}
    Total\ Storage = F \times N \times ((1 + \alpha) \times (B \times c))
\end{equation}

where:
\begin{itemize}
    \item \lbrack$F$\rbrack \space is the total number of files uploaded to the network.
    \item \lbrack$N$\rbrack  \space is the number of peers on the network.
    \item \lbrack$B$\rbrack  \space is the number of blocks the file was split into.
    \item \lbrack$\alpha$\rbrack  \space is the encoding parameter.
    \item \lbrack$c$\rbrack  \space is the constant number of bytes (40 bytes) for each map entry when mapping the 64-bit lattice ID to a 256-bit HASH.
\end{itemize}

%% file: inputs/app4.tex
\section{Performance Evaluations Insights}
\label{performance_insight}

We compare different metrics for single and collaborative repairs using depth=5, 7, and 10.
The figures show the average and standard deviation values of all data. 

\begin{figure}[htp]
	\centering
	\begin{subfigure}{0.65\linewidth}
		\includegraphics[width=\linewidth]{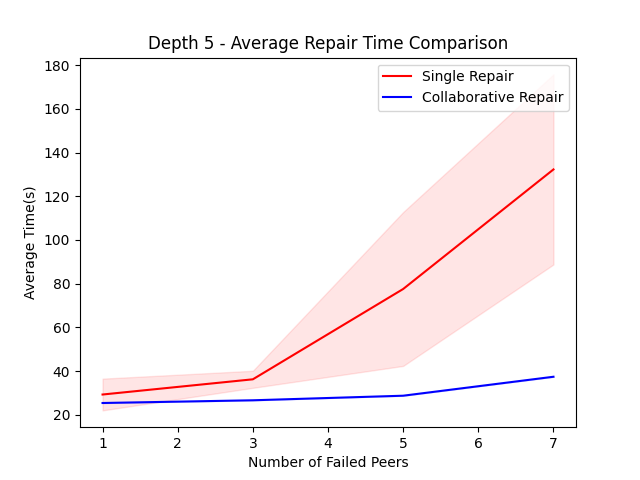}
		\caption{Depth 5}
		\label{total_time_recovery_A}
	\end{subfigure}
  \smallskip
	\begin{subfigure}{0.65\linewidth}
		\includegraphics[width=\linewidth]{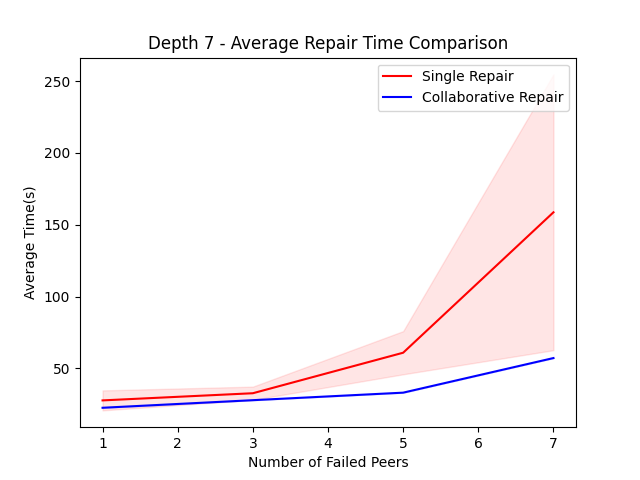}
		\caption{Depth 7}
		\label{total_time_recovery_B}
	\end{subfigure}
 \centering
	\begin{subfigure}{0.65\linewidth}
	        \includegraphics[width=\linewidth]{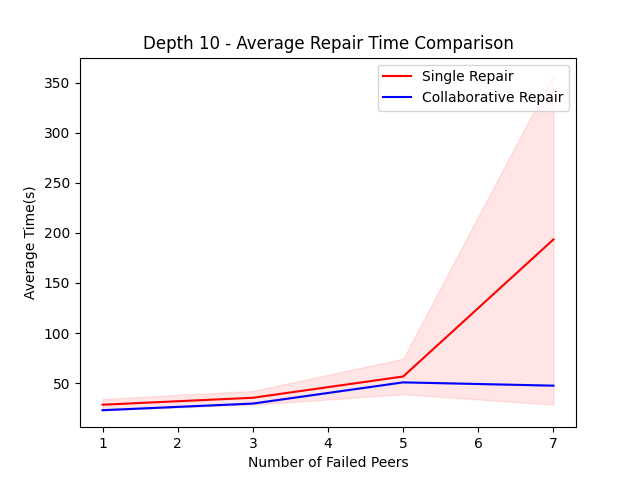}
	        \caption{Depth 10}
	        \label{total_time_recovery_C}
         \end{subfigure}
	\caption{Average total time to recover a file}
	\label{fig:total_time_recovery}
\end{figure}

\begin{figure}[ht]

	\centering
	\begin{subfigure}{0.65\linewidth}
		\includegraphics[width=\linewidth]{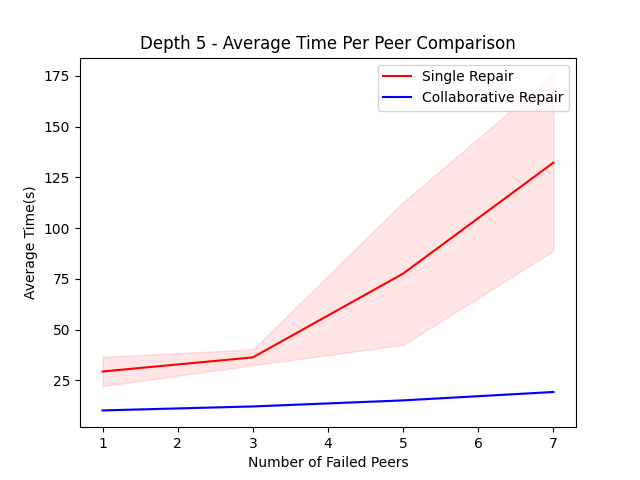}
		\caption{Depth 5}
		\label{total_time_recovery_peer_A}
	\end{subfigure}
 \smallskip
	\begin{subfigure}{0.65\linewidth}
		\includegraphics[width=\linewidth]{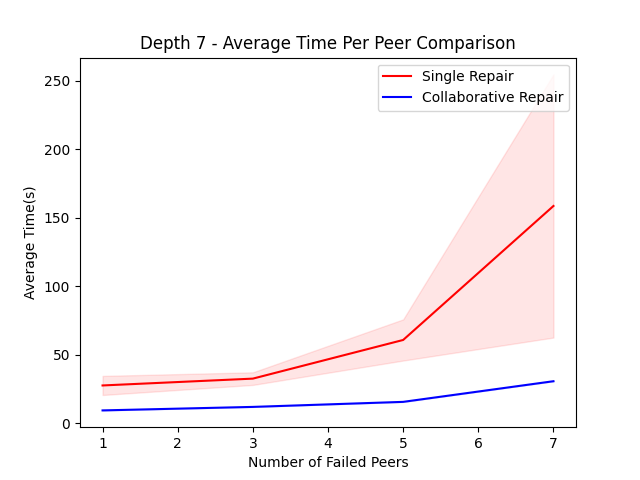}
		\caption{Depth 7}
		\label{total_time_recovery_peer_B}
	\end{subfigure}
 \smallskip
	\begin{subfigure}{0.65\linewidth}
	        \includegraphics[width=\linewidth]{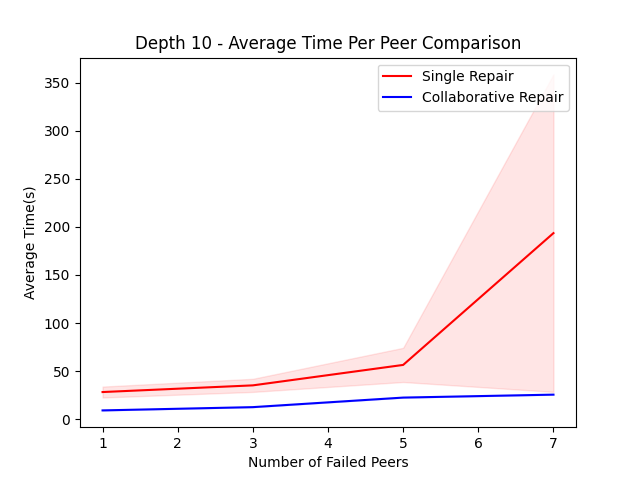}
	        \caption{Depth 10}
	        \label{total_time_recovery_peer_C}
         \end{subfigure}
	\caption{Average time per peer to recover a file}
	\label{fig:total_time_recovery_peer}
\end{figure}

\begin{figure}[ht]
	\centering
	\begin{subfigure}{0.65\linewidth}
		\includegraphics[width=\linewidth]{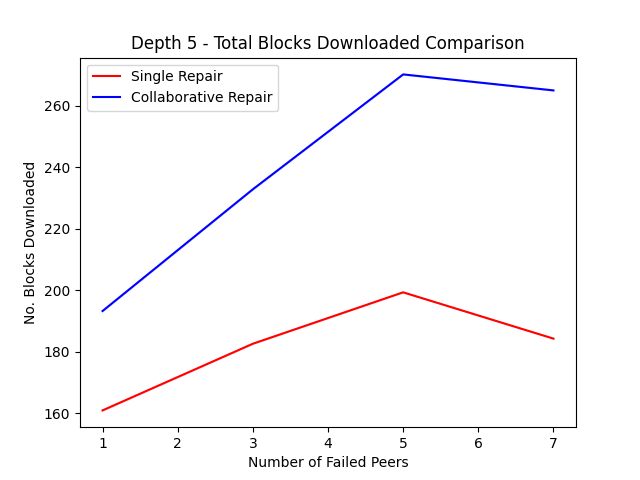}
		\caption{Depth 5}
		\label{total_blocks_download_A}
	\end{subfigure}
 \smallskip
	\begin{subfigure}{0.65\linewidth}
		\includegraphics[width=\linewidth]{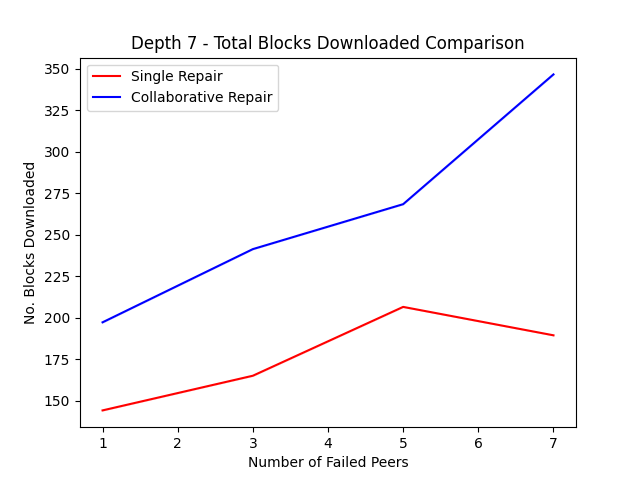}
		\caption{Depth 7}
		\label{total_blocks_download_B}
	\end{subfigure}
 \smallskip
	\begin{subfigure}{0.65\linewidth}
	        \includegraphics[width=\linewidth]{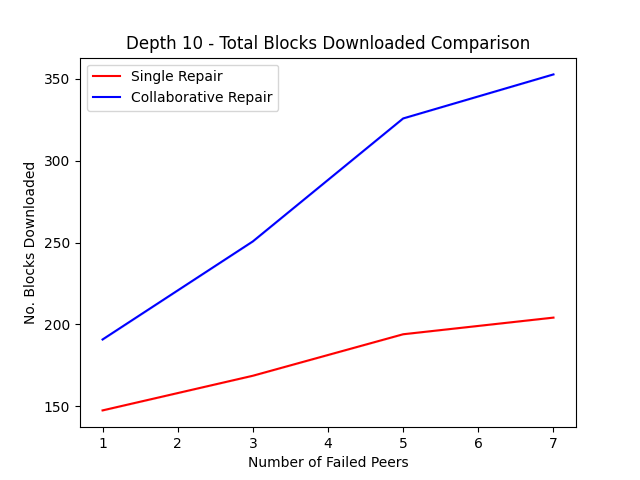}
	        \caption{Depth 10}
	        \label{total_blocks_download_C}
         \end{subfigure}
	\caption{Average total number of blocks downloaded}
	\label{fig:total_blocks_download}
\end{figure}

\begin{figure}[ht]
	\centering
	\begin{subfigure}{0.65\linewidth}
		\includegraphics[width=\linewidth]{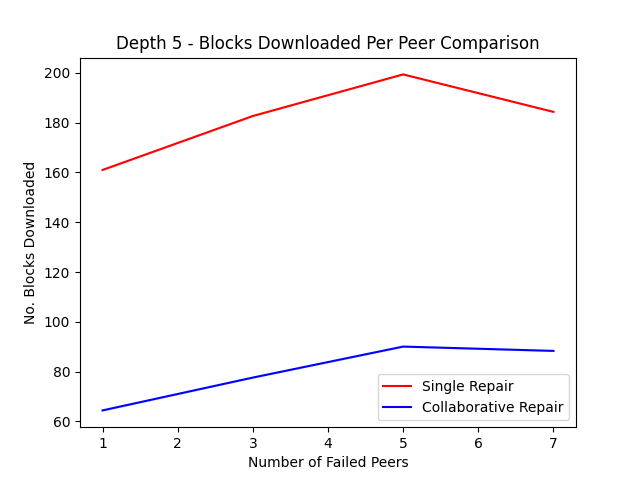}
		\caption{Depth 5}
		\label{total_blocks_download_peer_A}
	\end{subfigure}
 \smallskip
	\begin{subfigure}{0.65\linewidth}
		\includegraphics[width=\linewidth]{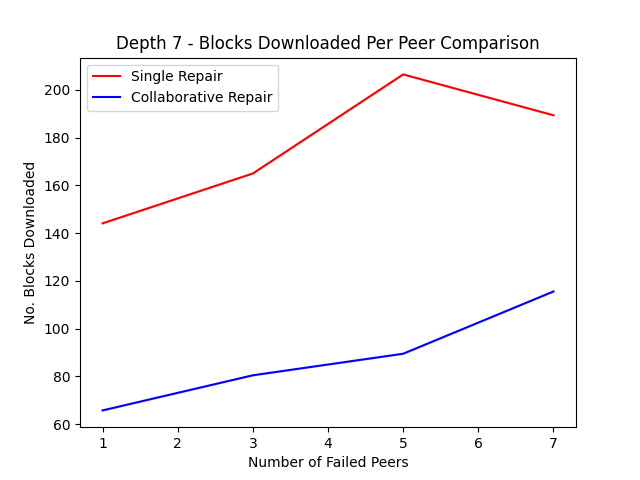}
		\caption{Depth 7}
		\label{total_blocks_download_peer_B}
	\end{subfigure}
 \smallskip
	\begin{subfigure}{0.65\linewidth}
	        \includegraphics[width=\linewidth]{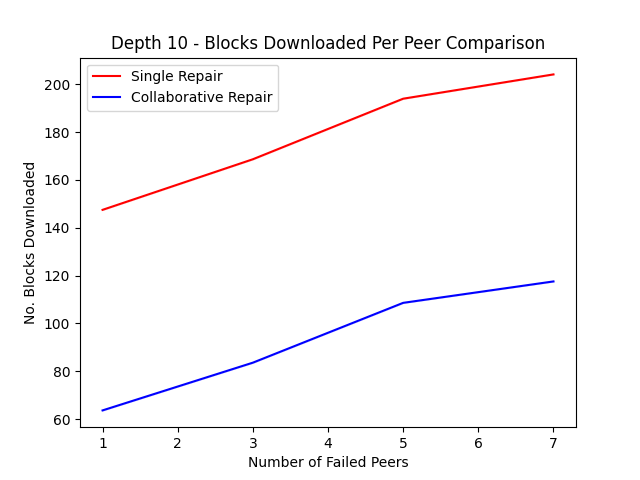}
	        \caption{Depth 10}
	        \label{total_blocks_download_peer_C}
         \end{subfigure}
	\caption{Average total number of blocks downloaded per peer}
	\label{fig:total_blocks_download_peer}
\end{figure}

%% file: main.bbl
\begin{thebibliography}{99}

\bibitem{ray2023web3}
Ray, Partha Pratim
\newblock {``Web3: A comprehensive review on background, technologies, applications, zero-trust architectures, challenges and future directions.''}
\newblock {\em Internet of Things and Cyber-Physical Systems} (2023)
\url{https://doi.org/10.1016/j.iotcps.2023.05.003}

\bibitem{wust2018you}
\newblock{W{\"u}st, Karl and Gervais, Arthur}
\newblock{``Do you need a blockchain?''}
\newblock{2018 IEEE Crypto Valley Conference on Blockchain Technology (CVCBT)}
\newblock{45--54}
\url{https://doi.org/10.1109/CVCBT.2018.00011}

\bibitem{tschorsch2022ipfs}
\newblock{Daniel, Erik and Tschorsch, Florian}
\newblock{``IPFS and friends: A qualitative comparison of next generation peer-to-peer data networks.''}
\newblock{2022 IEEE Communications Surveys \& Tutorials},
\newblock{24 1 31--52}
\url{https://doi.org/10.1109/COMST.2022.3143147}

\bibitem{ipfscluster}
\newblock{Protocol Labs}
\newblock{``IPFS Cluster - Automated data availability and redundancy on IPFS.''}
\url{https://ipfscluster.io/documentation}

\bibitem{benet2018filecoin}
\newblock{Benet, Juan and Greco, Nicola}
\newblock{``Filecoin: A decentralized storage network.''}
\newblock{Protocol Labs Whitepaper 2018}
\url{https://filecoin.io/filecoin.pdf}

\bibitem{benet2014ipfs}
\newblock{Benet, Juan}
\newblock{``IPFS-content addressed, versioned, p2p file system.''}
\newblock{Whitepaper 2014}
\url{https://doi.org/10.48550/arXiv.1407.3561}

\bibitem{maymounkov2002kademlia}
\newblock{Maymounkov, Petar and Mazieres, David}
\newblock{``Kademlia: A peer-to-peer information system based on the xor metric.''}
\newblock{IEEE International Workshop on Peer-to-Peer Systems 53--65 Springer 2002}
\url{https://doi.org/10.1007/3-540-45748-8_5}

\bibitem{buterin2014ethereum}
\newblock{Buterin, Vitalik and others}
\newblock{``Ethereum.''}
\newblock{Whitepaper 2014}
\url{https://ethereum.org/en/whitepaper}

\bibitem{jiang2021cryptokitties}
\newblock{Jiang, Xin-Jian and Liu, Xiao Fan}
\newblock{``Cryptokitties transaction network analysis: The rise and fall of the first blockchain game mania.''}
\newblock{Frontiers in Social Physics 2021 Volume 9 2021}
\url{https://doi.org/10.3389/fphy.2021.631665}

\bibitem{nygaard2021snarl}
\newblock{Nygaard, Racin and Estrada-Gali{\~n}anes, Vero and Meling, Hein}
\newblock{``Snarl: Entangled merkle trees for improved file availability and storage utilization.''}
\newblock{Proceedings of the 22nd International Middleware Conference 236--247 2021}
\url{https://doi.org/10.1145/3464298.3493397}

\bibitem{estrada2018alpha}
\newblock{Estrada-Gali{\~n}anes, Vero and Miller, Ethan and Felber, Pascal and P{\^a}ris, Jehan-Fran{\c{c}}ois}
\newblock{``Alpha entanglement codes: practical erasure codes to archive data in unreliable environments.''}
\newblock{2018 48th Annual IEEE/IFIP International Conference on Dependable Systems and Networks (DSN) 183--194}
\url{https://doi.org/10.1109/DSN.2018.00030}

\bibitem{ongaro2014search}
\newblock{Ongaro, Diego and Ousterhout, John}
\newblock{``In search of an understandable consensus algorithm.''}
\newblock{2014 USENIX Annual Technical conference (USENIX ATC 14) 305--319}
\url{https://www.usenix.org/conference/atc14/technical-sessions/presentation/ongaro}

\end{thebibliography}
